# REDUCING BACKGROUNDS IN THE HIGGS FACTORY MUON COLLIDER DETECTOR *†

N.V. Mokhov, S.I. Striganov#, I.S. Tropin

*Fermi National Accelerator Laboratory, Batavia IL 60510-5011, USA*

## ABSTRACT

A preliminary design of the 125 GeV Higgs Factory (HF) Muon Collider (MC) has identified an enormous background loads on the HF detector. This is related to the twelve times higher muon decay probability at HF compared to that previously studied for the 1.5-TeV MC. As a result of MARS15 optimization studies, it is shown that with a carefully designed protection system in the interaction region, in the machine-detector interface and inside the detector one can reduce the background rates to a manageable level similar to that achieved for the optimized 1.5-TeV case. The main characteristics of the HF detector background are presented for the configuration found..

*Work supported by Fermi Research Alliance, LLC under contract No. DE-AC02-07CH11359 with the U.S. Department of Energy through the DOE Muon Accelerator Program (MAP).

†Presented paper at the 5th International Particle Accelerator Conference, June 15-20, 2014, Dresden, Germany

#striganoov@fnal.gov

# REDUCING BACKGROUNDS IN THE HIGGS FACTORY MUON COLLIDER DETECTOR *

S.I. Striganov[#], N.V. Mokhov, I.S. Tropin, FNAL, Batavia, IL 60510, USA

*Abstract*

A preliminary design of the 125-GeV Higgs Factory (HF) Muon Collider (MC) has identified an enormous background loads on the HF detector. This is related to the twelve times higher muon decay probability at HF compared to that previously studied for the 1.5-TeV MC. As a result of MARS15 optimization studies, it is shown that with a carefully designed protection system in the interaction region, in the machine-detector interface and inside the detector one can reduce the background rates to a manageable level similar to that achieved for the optimized 1.5-TeV case. The main characteristics of the HF detector background are presented for the configuration found.

## INTRODUCTION

A Higgs Factory (HF) Muon Collider (MC) offers unique possibilities for studying the Higgs boson [1]. The impact of the radiation environment produced by muon decays is the fundamental and critical issue in determining the feasibility of HF and its detectors. Muon decays are identified as the major source of detector background at a MC [2-4]. The decay length for a 62.5 GeV muon is $3.9 \cdot 10^5$ m. With $2 \cdot 10^{12}$ muons per bunch, these result in $10^7$ decays per meter in a single pass. The HF ring considered here is designed [5] for 1000 to 2000 turns per a store with 30 stores per second. Electromagnetic showers induced by decay electrons in the collider components generate intensive fluxes of hadrons, muons, photons and daughter electrons, which create high background and radiation levels in the detector. This creates difficulties with reconstruction of tracks, deteriorates detector resolution and produces radiation damage in detector components.

## HIGGS FACTORY NOZZLE DESIGN

In comparison with a 1.5-TeV MC, the HF background problem is more complicated. There is 12 times more muon decays per meter. The coil apertures are much larger with magnets being shorter. The open region at the Interaction Point (IP) is about 4 times longer due to a longer bunch length. The nozzle tips are at ~28 cm from IP with a 5 cm radius at that location. The nozzle tapers down to an aperture of 2.5 cm at z=±49 cm to prevent direct hits of the beryllium beam pipe by decay electrons (Fig.1). Then the inner cone tapers to radius of 2 cm at z=±107 cm to reduce the number of high energy interactions near IP. After that, the nozzle aperture diameter equals to the 8σ beam envelope.

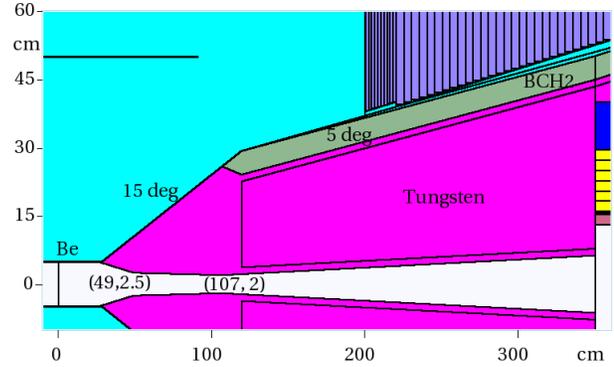

Figure 1: Tungsten nozzle.

The outer angle subtended by the nozzle in the region closest to IP (28-119 cm) is the most critical parameter: the larger angle provides the better background reduction, however, the impact on detector performance becomes higher. Two values of this angle were considered, 7.5 and 15 degrees. At z >119 cm, the angle is reduced to 5 degrees. Tungsten is encapsulated in a borated polyethylene shell to reduce the flux of low-energy neutrons.

## MARS MODELING OF BACKGROUND

Detector backgrounds are simulated with the MARS code [6,7]. A realistic 3D model of the entire HF collider ring was built with the SiD-like detector at IP [8]. A silicon vertex detector and a tracker are based on the design proposed for the CMS upgrade.

The background loads at HF MC for two nozzle angles are compared in Table 1 with results obtained in Ref. [4] for the 1.5-TeV MC.

Table 1: Number of particles entering detector per BX. ch. hadrons > 1MeV, γ and e± > 0.2 MeV, n > 0.1 MeV

| Particle | 750 GeV 10 deg | 62.5 GeV 7.5 deg | 62.5 GeV 15 deg |
|---|---|---|---|
| Neutrons | $4.1 \times 10^7$ | $3.2 \times 10^8$ | $5.2 \times 10^7$ |
| Ch. hadrons | $4.8 \times 10^4$ | $3.0 \times 10^6$ | $1.0 \times 10^4$ |
| e± | $1.0 \times 10^6$ | $3.0 \times 10^8$ | $2.0 \times 10^6$ |
| Photons | $1.8 \times 10^8$ | $8.1 \times 10^9$ | $2.8 \times 10^8$ |

The most optimal 15-degree nozzle design – that also includes a larger Be pipe at IP, additional shielding at the Machine-Detector Interface (MDI) along with tighter tungsten masks and liners inside the magnets [8] - reduces the background loads ~30 times for photons, ~150 times for electrons and positrons, and ~6 times for neutrons in comparison with the earlier 7.5-degree design. In the rest of the paper, results are presented for this optimal design.

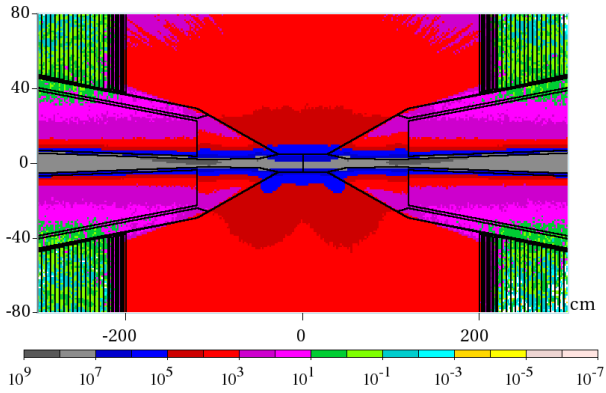

Figure 2. Gamma fluence (1/cm$^2$/BX).

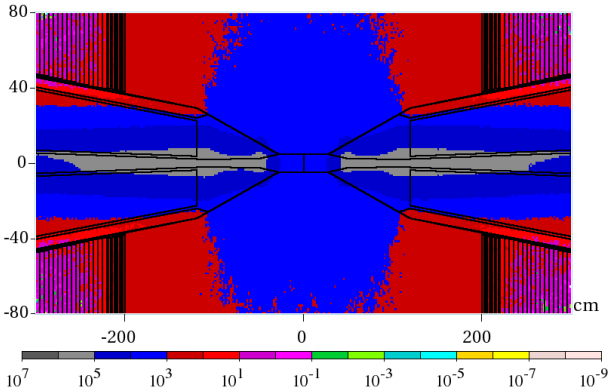

Figure 3. Neutron fluence (1/cm$^2$/BX).

Figs. 2 and 3 show – respectively - photon and neutron fluence isocontours in the central detector region. The photon fluence is noticeably larger than in the LHC detectors while for neutrons it is much lower [4].

About a half of the background loads (except muons) is produced by decay e$^\pm$ hitting the inner nozzle surface upstream of the minimal aperture at 107<|z|< 150 cm, 25% are from decay electron interactions with the nozzle jaws at 49<|z|<107 cm, ~22% is due to interactions with the jaws after IP, 30< |z| < 49 cm, and the rest is from e$^\pm$ interactions at |z| > 150 cm. Most of particles reaching the detector are result of backscattering from the nozzle aperture.

In MARS15 runs, a source term for detector simulations is generated for two 62.5 GeV muon beams with full information on the particle origin. It is for all particles entering the detector through the MDI surfaces. These surfaces are defined on the outer surface of the nozzle and beam pipe, and on the inner surface of the barrel and endcap electromagnetic calorimeters.

## BASIC FEATURES OF BACKGROUND

The origin of 98% of particles (except muons) entering the detector is muon decays at ±15 m from IP. As Fig. 4 shows, the muon decays inside the first three quadrupoles (±8 m) produce 70% of electron, positron, photon and hadron backgrounds while Bethe-Heitler muons hitting detector are created in the lattice as far as 40 m from IP.

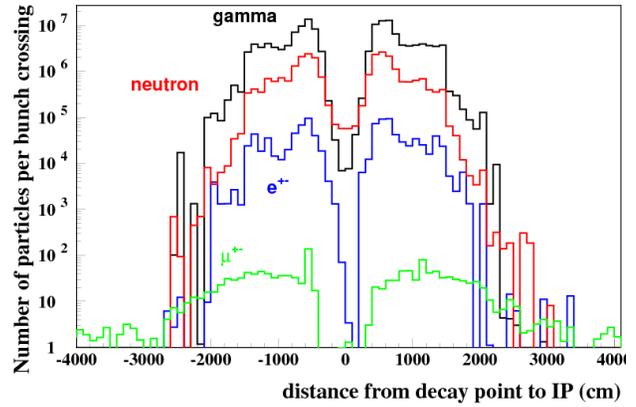

Figure 4. Number of background particles as a function of distance from IP to muon decay point along the beam.

The particle tracking cut-off energies were chosen to adequately understand the main features of the background balancing accuracy against CPU time. Using the MARS15 code with the EGS5 option, energy spectra of background e$^+$, e$^-$, photons and muons were simulated with a 10 keV threshold. It turns out that 99.3% of background photons and 97% of electrons have energy higher than 100 keV. Lowering the electron and photons energy cuts is very time consuming, but does not change background properties. A neutron component of the load is calculated down to 0.001 eV (Fig. 5). Mean kinetic energies are 45 MeV (positrons), 9 MeV (electrons), 7 MeV (photons), 1 MeV (neutrons), 3.6 GeV (muons) and 350 MeV (charged hadrons).

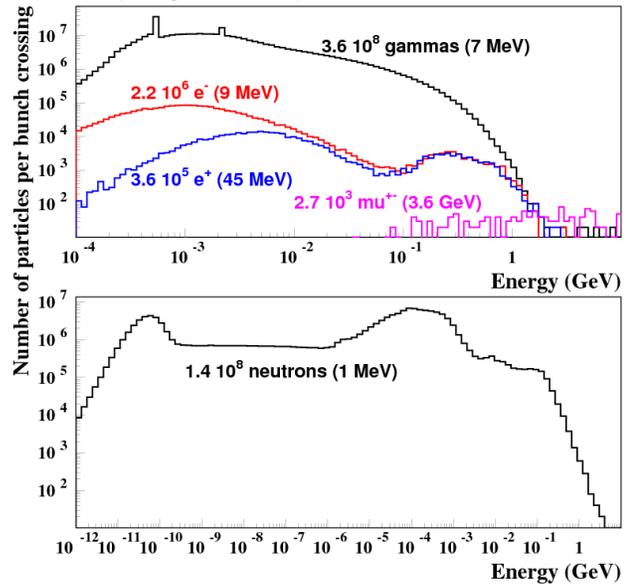

Figure 5. Energy spectra of background particles.

The time of flight (TOF) of background neutrons at the MDI surface has significant spread with respect to the bunch crossing as shown in Fig. 6. This property can help reduce neutron hit rates in detector using timing cuts. At the same time, photons, electrons and positrons at the MDI surface have a small lag relative to the incoming muon beam.

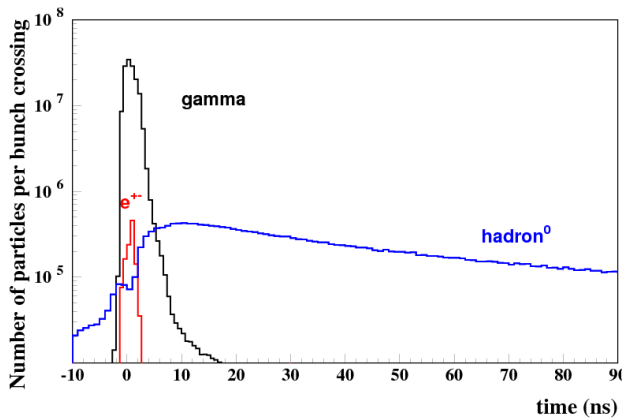

Figure 6. TOF distributions of background particles at MDI surface with respect to bunch crossing.

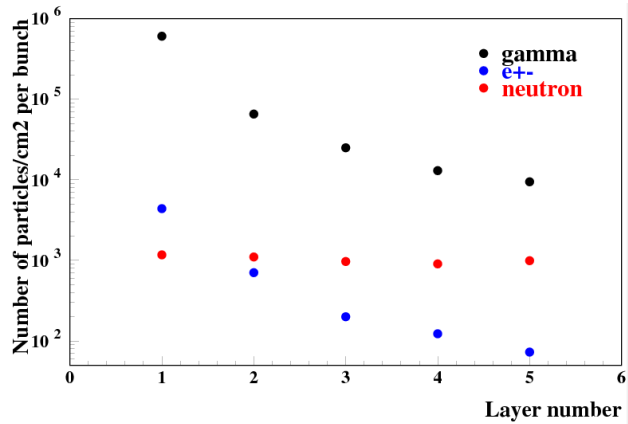

Figure 7. Fluxes of background particles in VXD.

The VXD barrel consists of five silicon layers with thickness of 75 μm. The radii of the layers range from 5.4 to 21.59 cm. The VXD barrel is very close to the unshielded beam pipe, therefore it is the hottest part of the system. The fluxes of background particles in the vertex detector are presented in Fig.7. It is seen that the number of photons and $e^{\pm}$ in the barrel decreases rapidly with the VXD layer radius. The neutron radial distribution in VXD is flat.

## VERTEX DETECTOR OCUPANCY

The calculated background source term at the MDI surfaces was used to simulate hit rates in the silicon VXD and tracker detectors. Backgrounds due to primary particles from the source, as well as secondary and tertiary particles produced in interactions of primaries with the IR materials are tracked in the detector magnetic field. The minimal cut-off energy for electrons, positrons and photons was reduced to 3 keV, for other particles it remains the same as in the source file.

The hit in a silicon detector volume is defined as a charged track which left the volume or stopped in it. The most of hits (90%) in VXD is produced by primary photons; primary neutrons are the origin of 6% hits, remaining 4% are created by primary electrons and positrons.

To calculate the detector occupancy, one needs to perform full simulation for the chosen detector geometry. At the same time, one can make an estimate of the hit rate using a low transport threshold. The electron range in silicon at 3 keV is 0.14μm. With this cut-off most of the secondary electrons are stopped in the same pixel where the outgoing/stopping charged track already produced hit. So, one can derive the number of hits from above.

The hit rate is maximal in the first VXD barrel layer where it reaches $4.6 \cdot 10^4$ hit/cm$^2$. A Fine Pixel CCD VXD was proposed for ILC in Ref. [9]. It has a very small pixel size of 5-10 μm and tolerates the occupancy up to 3%. If the first barrel layer consists of 5 μm pixels, the occupancy estimated from the above is 1.2%. The background loads in other vertex barrel and endcap layers are also acceptable even with a larger 20-μm pixel size. Reduction of hit rates in the detector was shown to be quite effective using various background rejection techniques [10] such as timing with < 100 ps time resolution in front-end ROC, double-layer geometry and others.

## CONCLUSION

A detector protection system based on the optimized tungsten nozzle and sophisticated MDI was developed for the Higgs Factory muon collider. This system allowed reduction of the background rates to a manageable level similar to that achieved for a 1.5-TeV MC. The main characteristics of particle backgrounds entering the collider detector were studied. The occupancy estimated for the silicon VXD barrel and endcap layers is quite acceptable for the modern detector technologies.